\documentclass[sigconf]{acmart}

\usepackage{graphicx}

\usepackage{amsmath}
\usepackage{xcolor}
\usepackage{wrapfig}
\usepackage{caption}

\usepackage{float}

\newcommand\mc{\multicolumn}

\copyrightyear{2024}
\acmYear{2024}
\setcopyright{rightsretained}
\acmConference[DAC '24]{61st ACM/IEEE Design Automation Conference}{June 23--27, 2024}{San Francisco, CA, USA}
\acmBooktitle{61st ACM/IEEE Design Automation Conference (DAC '24), June 23--27, 2024, San Francisco, CA, USA}
\acmDOI{10.1145/3649329.3657339}
\acmISBN{979-8-4007-0601-1/24/06}

\renewcommand{\bibliofont}{\fontsize{6.9pt}{7pt}\selectfont}

\begin{document}


\title{Deep Harmonic Finesse: Signal Separation in Wearable Systems with Limited Data}

\author{ Mahya Saffarpour$^{1}$, Kourosh Vali$^{1}$, Weitai Qian$^{1}$, Begum Kasap$^{1}$, Herman L. Hedriana$^{2}$, and Soheil Ghiasi$^{1}$} 
\thanks{
*This work was supported by the National Science Foundation (NSF) Grants 1838939, 1934568, and 1937158; National Center for Interventional Biophotonic Technologies (NCIBT) through NIH under Grant 1P41EB032840; and the University of California-Noyce Initiative}
\thanks{$^{1}$ The department of Electrical and Computer Engineering, UC Davis, One Shields Ave., Davis, CA, USA
        {(email of the corresponding author: \tt\small msaff@ucdavis.edu).}}%
\thanks{$^{2}$ The department of OB/GYN, UC Davis School of Medicine, Sacramento, CA, USA
        {}
\vspace{-7pt}
}%

\begin{abstract}

We present a method, referred to as Deep Harmonic Finesse (DHF), for separation of non-stationary quasi-periodic signals when limited data is available. The problem frequently arises in wearable systems in which, a combination of quasi-periodic physiological phenomena give rise to the sensed signal, and excessive data collection is prohibitive. Our approach utilizes prior knowledge of time-frequency patterns in the signals to mask and in-paint spectrograms. This is achieved through an application-inspired deep harmonic neural network coupled with an integrated pattern alignment component. The network’s structure embeds the implicit harmonic priors within the time-frequency domain, while the pattern-alignment method transforms the sensed signal, ensuring a strong alignment with the network. The effectiveness of the algorithm is demonstrated in the context of non-invasive fetal monitoring using both synthesized and \emph{in vivo} data. When applied to the synthesized data, our method exhibits significant improvements in signal-to-distortion ratio (26\% on average) and mean squared error (80\% on average), compared to the best competing method. When applied to \emph{in vivo} data captured in pregnant animal studies, our method improves the correlation error between estimated fetal blood oxygen saturation and the ground truth by 80.5\% compared to the state of the art.
\end{abstract}


\maketitle

\section{Introduction}

Wearable systems hold significant potential for improving users' wellness and quality of life. In some wearable systems, the collected sensor data is impacted by multiple quasi-periodic non-stationary artifacts that may be considerably stronger than the signal of interest. Examples arise in tissue sensing applications, such as tissue oximetry or blood glucose monitoring, in which the sensor output signal is influenced by a parameter of interest as well as other quasi-periodic physiological phenomena, such as heart rate, respiration and Mayer waves. As obtaining large amount of high-quality data is expensive, and often impractical, successful development of such applications relies on isolation of the target signal from the sensed signal using limited available data.

In this paper, we present a method to separate quasi-periodic mixed-signals in the time-frequency space using only a single input data sample under the assumption that 1) the signals are quasi-periodic and non-stationary; 2) only a single detector measurement is available; and 3) signal fundamental frequencies, but not their amplitudes, are known either through auxiliary sensing modalities or preliminary analysis of the mixed signal, e.g.,  \cite{niedzwiecki2005estimation, zhang2016fundamental, KUBAI}. Notably, this method addresses two major challenges: 1) the potential overlap of signal frequencies, which renders classic frequency-based filtering techniques ineffective; and 2) the limitation of having only a single sample, which precludes the use of traditional machine learning methods that require a training dataset.

The proposed method, named Deep Harmonic Finesse (DHF), separates signals in iterations. In each iteration, the problem is approached as a process of masking selected regions of the mixed signal's spectral representation, and in-painting them using a deep prior structure inspired by \emph{Deep Prior} \cite{deepimageprior, deepaudioprior_harmonicconv}. The masking step leverages the known frequency information of the source signals to conceal information of all, but one of the signal sources. The algorithm then in-paints the masked parts of the spectrogram to eliminate all masked signal sources from the mix and recover the target signal values. The recovered signal is removed from the mix, and the process is iteratively applied to the residual signal. The proposed structural implicit priors comprise two key components: a signal pattern-aligner and a specialized harmonic convolutional neural network. The pattern-aligner resamples and transforms the input signal into a new space, providing the network with an input that aligns well with its expected deep prior pattern.

To validate the effectiveness of our proposed method, we apply it on both synthesized  data, and \emph{in vivo} data captured in animal studies using a wearable transabdominal fetal pulse oximetery (TFO) device. This device senses a noisy photoplethysmography (PPG) signal influenced by three quasi-periodic dynamics: respiration, maternal pulsation, and fetal pulsation. \cite{fong2020design}. Compared to the best competing algorithm applied to the synthesized data, our proposed method improves signal-to-distortion ratio and mean squared error of extracted signals by an average of 26\% and 80\%, respectively. Compared to the prior work applied to the \emph{in vivo} data, our method improves the correlation between estimated fetal blood oxygen saturation and the ground truth by 80.5\%.

\section{Related Work}
Single-detector signal separation methods can be broadly classified into three categories: (1) Analytical component decomposition methods with or without consideration of the periodic behavior, (2) Deep learning methods that require ample datasets, and (3) Deep prior learning methods.

\textbf{Analytical methods:} Empirical Mode Decomposition (EMD) \cite{EMD} and Variational Mode Decomposition (VMD) \cite{VMD} decompose a time-domain signal into Intrinsic Mode Functions (IMFs), representing different sources. EMD considers signal's local extremas for extracting IMFs while VMD defines a variational problem assuming each source to have a compact frequency representation around a central frequency which tends to provide better frequency separation than EMD. Non-negative Matrix Factorization (NMF) \cite{NMF} operates in both time and frequency domains. In time-frequency, it breaks down the 2-D matrix into two product matrices, which can be interpreted as distinct sources.


Presented for sound signals' separation in time-frequency domain, REpeating Pattern Extraction Technique (REPET) \cite{REPET} works under the assumption that a sound signal can be represented as sum of a repeating background with a varying foreground. An extension of REPET, REPET-Extended, can adapt to changes in the repeating structure over time, providing better separation quality when the repeating parts of the signal are non-stationary.

\textbf{Deep learning methods:} Due to availability of large datasets in audio applications, a vast body of research has studied deep learning audio source separation methods \cite{narayanaswamy2020unsupervised}\cite{aud2}\cite{aud3}\cite{aud4}, which are not applicable when only limited training data is available.

\textbf{Deep Priors:} In their pioneering work, Ulyanov et al. demonstrated that the structure of a generative convolutional deep neural network, $f_{\theta}$, can be used as a prior for image denoising, in-painting and super-resolution tasks using only one sample in the dataset \cite{deepimageprior}. The network receives a random vector, $z$, as input, and is subsequently trained to generate an output that is \emph{similar} to a \emph{single}, given noisy, low-quality or masked image, $x_0$. More formally, the training minimizes the cost function $min_{\theta}E(f_{\theta}(z);x_0)$, where $E$ is a task specific term.


Zhang et al. adopted this idea for audio processing by designing harmonic convolutions, and showed how time-frequency priors emerge from such neural networks \cite{deepaudioprior_harmonicconv}. They argue that the natural statistics of images and audio signals are different than images, and therefore, their prior design should differ. They use the patterns of harmonics in time-frequency plain to redefine neighborhood in frequency in their proposed deformable harmonic convolution. The convolution operation between the noisy time-frequency spectrogram, $X[\omega,\tau]$, and the kernel $K$ with $H$ total harmonics at any specific frequency location, $\hat{\omega}$, and time, $\hat{t}$, is formulated in equation \ref{eq:conv}. By this definition, the convolution output at each frequency is the weighted sum of input at integer multiples of that specific frequency. Equation \ref{eq:conv2} extends this definition to consider backward harmonic access by introduction of an anchor, $n$, to the equation.
\begin{align}
    \vspace{-5pt}
    \label{eq:conv}
    &(X*K)[\hat{\omega},\hat{t}]=\sum_{k=1}^{H}\sum_{t=-T}^{T} X[k\hat{\omega},\hat{t}-t]K[k,t]\\
    \label{eq:conv2}
    &(X*K)[\hat{\omega},\hat{t}]=\sum_{k=1}^{H}\sum_{t=-T}^{T} X[\frac{k\hat{\omega}}{n},\hat{t}-t]K[k,t]
\end{align}

Using a single-detector sound in time-frequency space, \cite{deepaudioprior_separation} uses the auditory coherence and discontinuity feature of sound sources to separate the source signals using a deep convolutional prior technique. The technique utilizes two separate parallel networks per source that generate the signal and its silencer mask, mixing them together at their output to compare against the mixed input signal. However, this method is ineffective for separation of continuous quasi-periodic signals that inherently do not have silence gaps.

\begin{figure*} [t]

\vspace{-5pt}
\centering
  \includegraphics[width=7in]{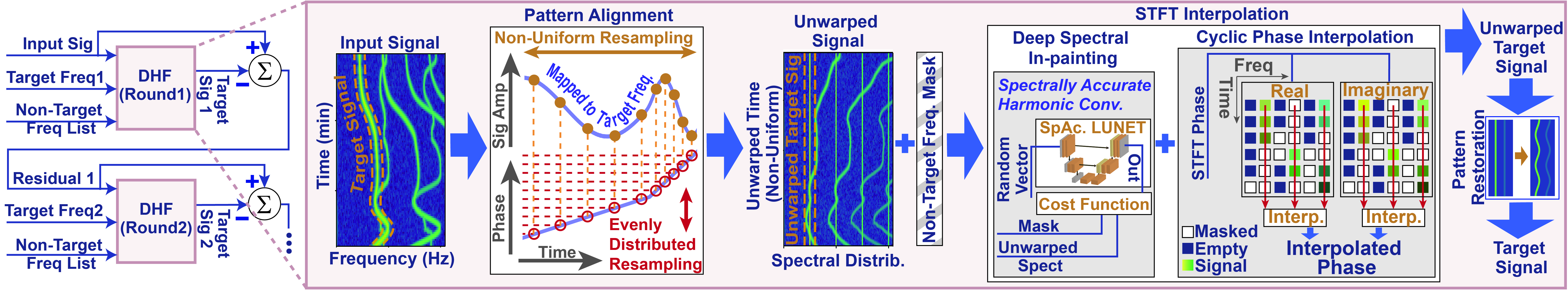}
  \caption{Overview of the proposed signal separation and DHF procedure.}
  \label{fig:procedure}

\vspace{-5pt}
\end{figure*}

\section{Proposed Methodology}
We propose an iterative method for signal separation in multi-source quasi-periodic signals using only a single mixed-signal in the dataset, named DHF. Each round begins by selecting the signal to be separated, followed by a proposed deep prior learning procedure for masking and in-painting the Short-Time Fourier Transform (STFT). The mask conceals all significant spectral components of sources other than the selected one. In-painting interpolates the masked regions of the selected signal, targeting amplitude and phase separately, to recover values during spectral overlaps and crossover. The structural prior used for spectrogram in-painting consists of two complementary components: a signal pattern aligner and a specialized convolutional neural network that we call the Spectrally Accurate Light U-Net (SpAc LU-Net). The pattern aligner resamples and unwarps the mixed signal, transforming it into a space where the selected source is strictly periodic (holding a steady fundamental frequency of 1 Hz). Given the strict periodic pattern over time and the harmonic patterns in spectrum, SpAc LU-Net defines the neighborhood in time and frequency by dilation and accessing integral multiples of target bin, respectively. Concurrently, phase interpolation is performed by separately interpolating the real and imaginary parts of each frequency bin's phase over time, and then recalculating the interpolated phase. The spectrogram in-painting results, joined with the interpolated phase information, are then passed through the inverse short time Fourier transform (ISTFT) and transformed back to the original space (referred to as pattern restoration) to deduce the separated signal time-serie.  After subtracting the separated source from the mixed signal, the residual is used for subsequent rounds of signal separation. Figure \ref{fig:procedure} sketches an overview of the proposed method, where each DHF block corresponds to one round of signal separation.

\subsection{Target Pattern Alignment}

In each round of signal separation, specific to a chosen target source, we preprocess the mixed signal, $X$, and unwarp it with respect to the fundamental frequency of the target source, $f_{ts}[n]$, locking a constant fundamental frequency of 1 Hz for that particular source in the unwarped signal, $X'$.

The original mixed signal space is defined in Equation \ref{eq:warped1}. With a full period of the quasi-periodic target source showing a phase change from $0$ to $2\pi$, Equation \ref{eq:warped2} computes the unrolled phase at each sampling time, $t[n]$.

\begin{align}
 \vspace{-5pt}
    \label{eq:warped1}
    &(t[n], X[n]),\;\; where\;\; t[n]=n\Delta t=\frac{n}{F_s}\\
    \label{eq:warped2}
    &\Phi [n]=2 \pi \sum_{i=1}^n f_{ts}[i] \Delta t=\frac{2 \pi}{F_{s0}} \sum_{i=1}^n f_{ts}[i]
\end{align}

Fluctuations in the fundamental frequency result in a variable phase interval between successive samples. To unwarp to a constant fundamental frequency of $1 Hz$, the new phase intervals between every pair of consecutive samples of $X'$ must remain constant, as indicated in Equation \ref{eq:unwarping1}. The unwarped signal $X'$ is calculated by two sequential interpolations, detailed in Equations \ref{eq:unwarping2}-\ref{eq:unwarping3}, first finding the timestamps of the evenly distributed phase intervals, $t'[m]$, and then the mixed signal values at those timestamps, $X'[m]$.

\begin{align}
 \vspace{-5pt}
    \label{eq:unwarping1}
    &\forall m, \Phi'[m]-\Phi'[m-1]=\frac{2\pi}{F'_s}\\
    \label{eq:unwarping2}
    &(\Phi[n] , t[n]) \to (\Phi'[m], t'[m])
\end{align}

\begin{align}
 \vspace{-5pt}
    \label{eq:unwarping3}
    &(t[n], X[n])\to(t'[m],X'[m])
\end{align}

\subsection{Deep Prior Architecture}

The pattern-aligned time-frequency spectrogram of quasi-periodic signals encapsulates critical features that need to be considered when constructing effective priors for the task of signal separation. These features outline the spatial neighborhood in both the time and spectral dimensions. Locking a constant fundamental frequency for the target source suggests that, to access preceding and subsequent amplitudes of that source over time, the convolutional kernel should exclusively examine the same frequency bin via dilation in time. Neighbors in frequency are defined as integral multiples of the current frequency, representing the harmonics, as opposed to the immediate next pixel (as in convolutional kernels). Equation \ref{eq:newconv} extends equation \ref{eq:conv} to add time dilation for constant frequency access in time.

\begin{align}
 \vspace{-5pt}
    \label{eq:newconv}
    &(X*K)[\hat{\omega},\hat{t}]=\sum_{k=1}^{H}\sum_{t=-T}^{T} X[k\hat{\omega},\hat{t}-D_{conv}t]K[k,t]
\end{align}

\begin{figure}[t]
\vspace{-5pt}
 \hfill\begin{minipage}{.47\textwidth}\centering
  \includegraphics[width=.38\paperwidth]{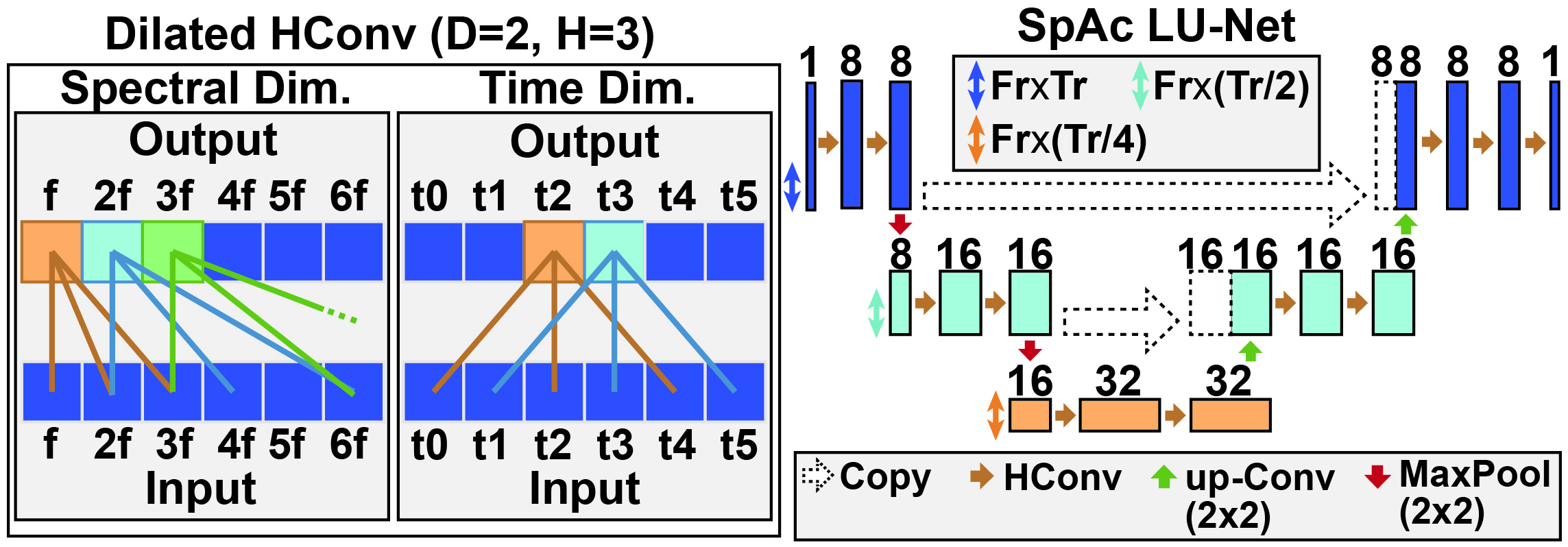}
  \caption{Dilated harmonic convolution and the structure of SpAc LU-Net.}
  \label{fig:conv}
 \end{minipage}
 \vspace{-5pt}
\end{figure}

We have adapted the U-Net architecture \cite{unet} as the foundation of our neural structure while substituting standard convolution kernels with dilated harmonic convolutions. Moreover, two design principles are maintained for harmonic integrity. Firstly, pooling in the frequency dimension is prohibited, ensuring the size of this dimension remains unchanged throughout the network. Secondly, only forward integral multiples are regarded as harmonic neighbors to guarantee spectral accuracy at all frequencies. Figure \ref{fig:conv} depicts the proposed SpAc LU-NET structure and the function of the proposed dilated harmonic convolution in time and frequency dimensions.

\subsection{Signal Separation by In-Painting}

We use masking and in-painting to separate sources in quasi-periodic mixed signals.  At each round, we use a binary mask to conceal all significant harmonics of non-targeting sources from the cost function. The in-painting cost function is similar to the original proposition by \cite{deepimageprior} for image in-painting applications. The selective visibility of the mask allows the optimizer to focus solely on the target source, in-painting the missing cross-overs and overlaps to retrieve the desired source. Equation \ref{eq:cost} presents the in-painting cost function, wherein $S_{mixed}$ and $S_{out}$ represent the mixed spectrogram and the neural network's output spectrogram, respectively.

\begin{figure*} [t]
\vspace{-5pt}
\centering
  \includegraphics[width=7.2in]{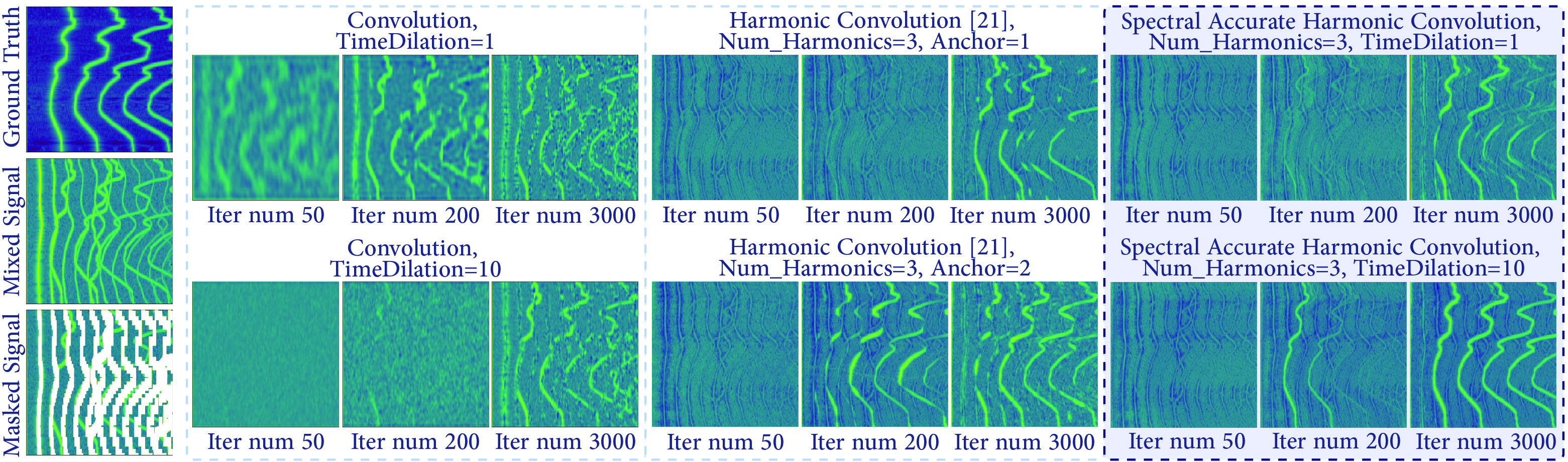}
  \caption{Example in-painting performance comparison of different convolutional kernels with the same network structure.}
  \label{fig:examplerun}

\vspace{-5pt}
\end{figure*}

\begin{align}
 \vspace{-5pt}
    \label{eq:cost}
    E(S_{out};S_{mixed})=||mask*(S_{out}-S_{mixed})||^2
\end{align}

Figure \ref{fig:examplerun} investigates the effectiveness of the proposed neural network structure, SpAc LU-NET with harmonic convolutions, in its role as implicit priors for learning quasi-periodic time-frequency patterns. It provides a comparative analysis of learning outcomes from different convolution kernels and network configurations receiving the same masked image. A comparison between conventional convolution kernels and harmonic convolution layers reveals the repetitive vertical frequency patterns, which are noticeable even at the initial stages of learning. However, the baseline harmonic convolution setup in \cite{deepaudioprior_harmonicconv} uses anchors larger than one (permits inaccurate backward harmonic neighboring) and max-pooling in frequency (results in inaccurate spectral representation). This setup weakens the prior structure and increases the noise. The Spectrally Accurate design, which builds on the previous setup by maintaining the anchor at 1 and eliminating frequency folding, shows a notable reduction of noise especially when using the time dilation that aligns well with the constant-frequency patterns after the unwarping step.

\subsection{Cyclic Phase Interpolation}

Phase interpolation of the STFT data is conducted independently from the spectrogram in-painting. Initially, we extract the phase information for each signal separation round from the unwarped STFT complex values. Utilizing the previously generated mask, we then estimate the phase during overlaps and crossovers through interpolation. Owing to the strictly periodic patterns in time-frequency space achieved after pattern alignment, we interpolate each frequency bin over time individually, yet concurrently with others. For each bin, we interpolate both the real and imaginary components of the phase, subsequently calculating the interpolated phase by integrating these results. This is to ensure the cyclic nature of the phase is preserved during interpolation.


\setlength{\arrayrulewidth}{.1em}
\begin{table}[t]
\vspace{-5pt}
\caption{Overview of the synthesized Mixed signals.}
\label{table:datasets}
\begin{center}
\setlength{\tabcolsep}{2pt}
\begin{tabular}{@{\extracolsep{4pt}}llccccc@{}}
    \mc{2}{l}{}&\mc{1}{c}{Syn.}&\mc{1}{c}{Syn.}&\mc{1}{c}{Syn.}&\mc{1}{c}{Syn.}&\mc{1}{c}{Syn.}\\
    \mc{2}{l}{}&\mc{1}{c}{MSig1}&\mc{1}{c}{MSig2}&\mc{1}{c}{MSig3}&\mc{1}{c}{MSig4}&\mc{1}{c}{MSig5}\\

    \hline \\[-2ex]
    source1 & mean(A) & 0.08 & 0.08 & 0.4 & 0.74 & 0.6 \\
         & std(A)  & 0.02 & 0.01 & 0.1 & 0.1 & 0.2 \\
         & $f_{min}$ & 0.9  & 0.8  & 1.4 & 0.5 & 0.5 \\
         & $f_{max}$ & 1.7  & 1.2  & 2.3 & 0.9 & 0.9 \\
    \hline \\[-2ex]
    source2 & mean(A) & 0.03 & 0.06 & 0.03 & 0.08 & 0.07\\
            & std(A)  & 0.01 & 0.02 & 0.01 & 0.01 & 0.01\\
            & $f_{min}$ & 1.8  & 1.0 & 1.6   & 1.1 & 1.0 \\
            & $f_{max}$ & 3.0  & 2.1 & 3.0   & 1.8 & 2.0 \\
    \hline \\[-2ex]
    source3 & mean(A) & -- & -- & -- & 0.06 & 0.04 \\
         & std(A)     & -- & -- & -- & 0.01 & 0.01 \\
         & $f_{min}$    & -- & -- & -- & 1.8 & 2.1 \\
         & $f_{max}$    & -- & -- & -- & 2.9 & 3.5 \\
    \hline \\[-2ex]
    noise  & mean & 0.0 & 0.0 & 0.0 & 0.0 & 0.0 \\
           & std & 0.003 & 0.01 & 0.04 & 0.01 & 0.001 \\

    \hline
\end{tabular}
\end{center}
\vspace{-5pt}
\end{table}

\section{Experimental Results}

\subsection{Synthesized Data}

\begin{figure}[t]
\vspace{-5pt}
\hfill\begin{minipage}{.5\textwidth}\centering
  \includegraphics[width=.39\paperwidth]{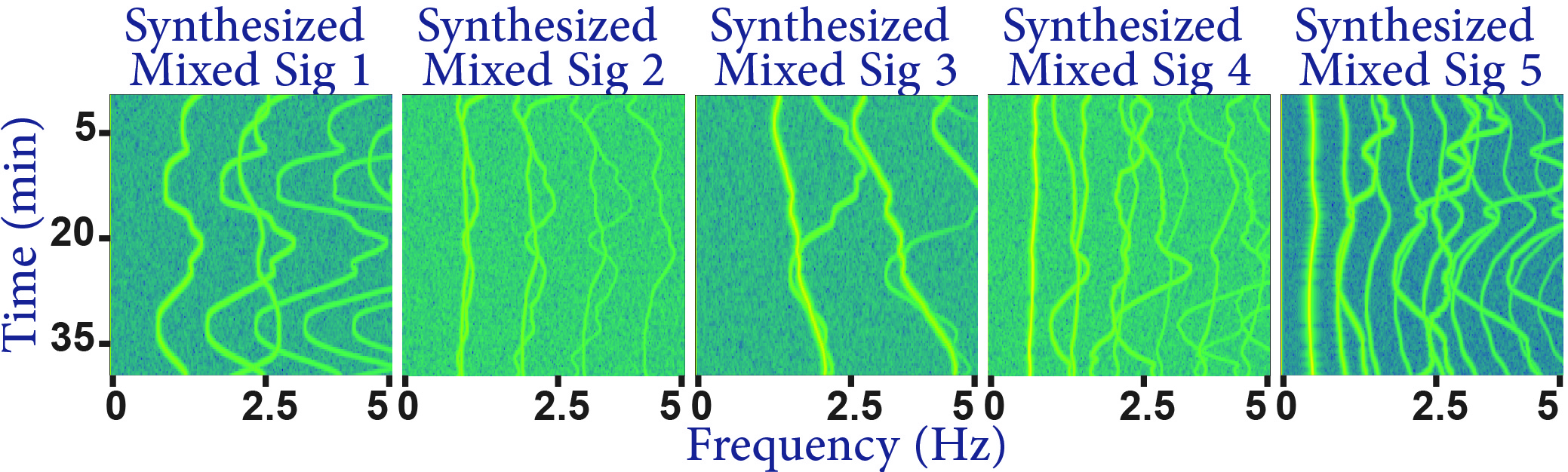}
  \caption{Time-frequency spectrogram of the synthesized mixed signals dataset generated for signal separation.}
  \label{fig:datasets}
 \end{minipage}
 \vspace{-5pt}
\end{figure}

We have created a tool for generating synthesized quasi-periodic timeseries, characterized by the desired input function per period, time duration per period list, and amplitude per period list. We have generated 5 distinct synthesized mixed quasi-periodic signals with sampling frequency of $100$ for TFO application. Each mixed signal has 2-3 sources for respiration, maternal pulsation, and fetal pulsation. The respiration PPG shape is extracted from real sheep experiment after filtering out other dynamics \cite{fong2020design, vali2021estimation}. The pulsation PPG shape is randomly extracted from MIMIC-IV dataset \cite{mimic_iv, physionet}. The spectrogram of the generated mixed signals is depicted in Figure \ref{fig:datasets}. The details of the synthesized signals are further explained in Table \ref{table:datasets}. Mixed Signals 1-3 each have two sources (maternal pulsation and fetal pulsation). The sources in  Mixed Signal 1 show interference in the second harmonic of the target source, while  Mixed Signal 2 has interference on the first harmonic. In  Mixed Signal 3 the second source has less than $\times0.1$ of the dominant source's amplitude. Mixed Signals 4-5 have three signals each (respiration, maternal pulsation, and fetal pulsation), with differences in overlap duration and complexity, and less than $\times0.1$ of amplitude on their third source compared to their first source.

\subsection{Signal Separation Results: Synthesized Data}

\begin{figure}[t]
\vspace{-5pt}
 \hfill\begin{minipage}{.5\textwidth}\centering
  \includegraphics[width=.38\paperwidth]{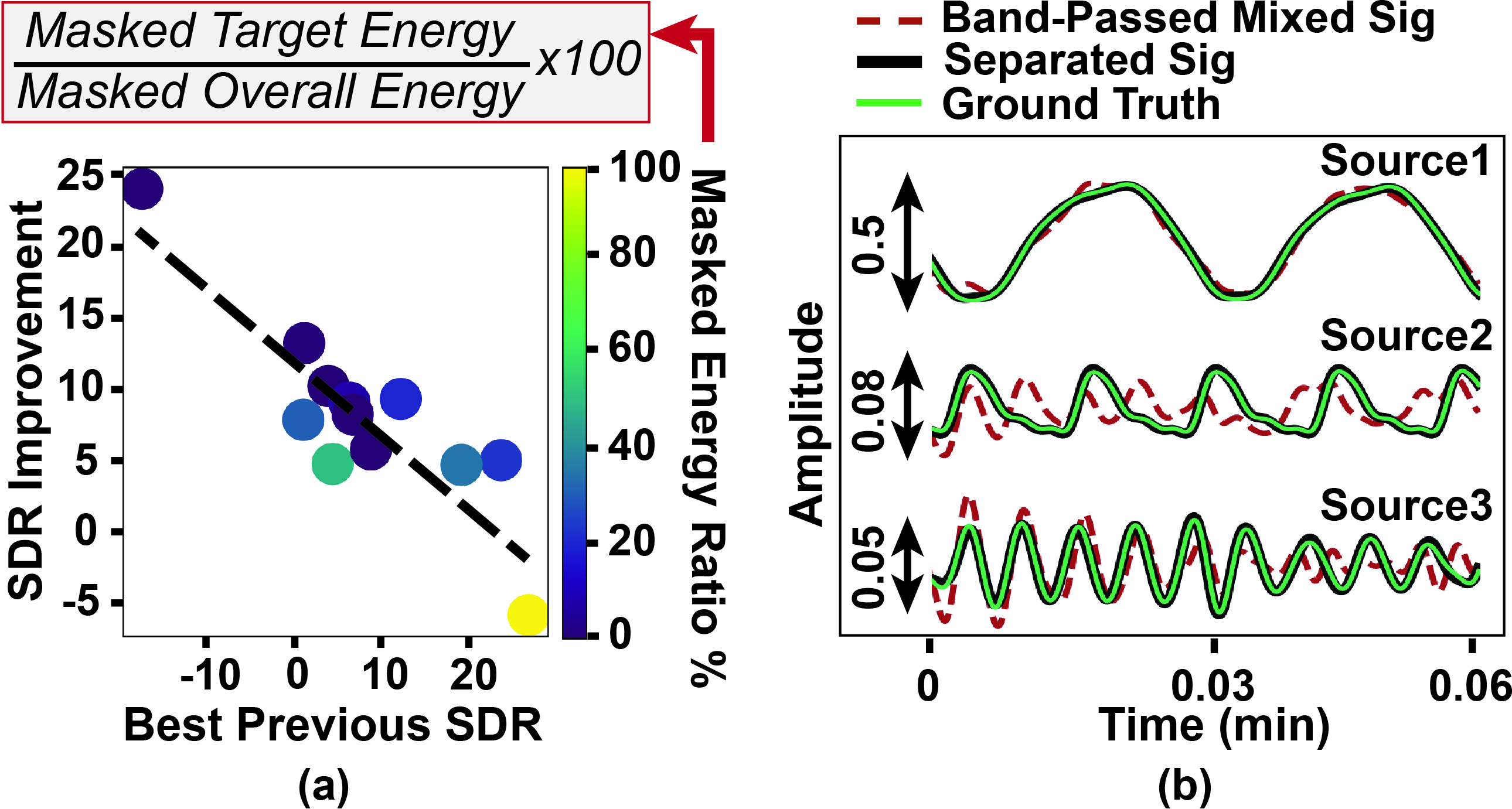}
  \caption{a) SDR comparison versus best previous method showing notable improvement in more difficult cases. b) Synthesized signal separation example by DHF.}
  \label{fig:wavesep}
 \end{minipage}
 \vspace{-5pt}
\end{figure}

Figure \ref{fig:wavesep} displays an example of signal separation results when DHF is applied to the mixed signal 5. The waveform shows a very close decomposition for all three sources. We further explore the statistics of DHF performance in comparison to previous signal separation methods when applied on the generated mixed signals.

\setlength{\arrayrulewidth}{.1em}
\begin{table*}[tb]
\vspace{-5pt}
\caption{Performance comparison of various signal separation methods applied on synthesized mixed signals 1-5. Best performance per source separation is highlighted.}
\label{table:finalres}
\begin{center}
\setlength{\tabcolsep}{1pt}
\resizebox{\textwidth}{!}{
\begin{tabular}{@{\extracolsep{2pt}}llcccccccccccccc@{}}
    \mc{2}{l}{}&\mc{2}{c}{EMD \cite{EMD}}&\mc{2}{c}{VMD \cite{VMD}}&\mc{2}{c}{NMF \cite{NMF}}&\mc{2}{c}{REPET \cite{REPET}}&\mc{2}{c}{REPET-Ext. \cite{REPET}}&\mc{2}{c}{Spect. Masking \cite{spectralmasking}}&\mc{2}{c}{DHF}\\

    \cline{3-4} \cline{5-6} \cline{7-8} \cline{9-10} \cline{11-12} \cline{13-14} \cline{15-16} \\[-1.5ex]

    \mc{2}{l}{}&\mc{1}{c}{SDR(db)}&\mc{1}{c}{MSE}&\mc{1}{c}{SDR(db)}&\mc{1}{c}{MSE}&\mc{1}{c}{SDR(db)}&\mc{1}{c}{MSE}&\mc{1}{c}{SDR(db)}&\mc{1}{c}{MSE}&\mc{1}{c}{SDR(db)}&\mc{1}{c}{MSE}&\mc{1}{c}{SDR(db)}&\mc{1}{c}{MSE}&\mc{1}{c}{SDR(db)}&\mc{1}{c}{MSE}\\

    \hline \\[-2ex]
    Syn. MSig1   & source1 & -1.38 & 7.4e-4 & 7.32 & 1.5e-4 & -9.03 & 8.9e-4 & 4.68 & 2.0e-4 & 9.91 & 1.0e-4 & 12.31 & 6.4e-5 & \textbf{21.63} & \textbf{7.4e-6} \\
        & source2 & -6.17 & 1.3e-4 & 3.17 & 1.1e-4 & -7.53 & 1.3e-4 & -0.77 & 6.4e-05 & -10.82 & 1.1e-4 & 6.44 & 3.3e-5 & \textbf{15.51} &\textbf{4.1e-6} \\
    \hline \\[-2ex]
    Syn. MSig2   & source1 & -6.36 & 9.1e-4 & 3.14 & 7.1e-4 & -4.58 & 7.8e-4 & 0.09 & 4.8e-4 & 4.82 & 3.4e-4 & 4.51 & 3.5e-4 & \textbf{9.29} &\textbf{1.1e-4}  \\
            & source2 & -21.75 & 7.2e-4 & -21.06 & 7.0e-4 & -4.98 & 6.4e-4 & -1.25 & 4.5e-4 & -6.2 & 4.4e-4 & 1.16 & 5.6e-4 & \textbf{9.02} & \textbf{9.2e-5} \\
    \hline \\[-2ex]
    Syn. MSig3   & source1 & 5.65 & 5.3e-3 & 7.24 & 3.9e-3 & -8.79 & 2.2e-2 & 6.59 & 3.3e-3 & 14.36 & 8.1e-4 & \textbf{26.95} & \textbf{5.7e-5} & 21.18 & 2.1e-4  \\
            & source2 & 0.07 & 2.6e-4 & -0.15 & 1.8e-4 & -0.18 & 8.3e-4 & -0.04 & 2.7e-4 & -1.63 & 2.1e-4 & -17.3 & 9.9e-3 & \textbf{6.96} & \textbf{4.0e-5} \\
    \hline \\[-2ex]
    Syn. MSig4   & source1 & 5.2 & 1.1e-2 & 15.16 & 1.5e-3 & -4.95 & 3.6e-2 & 3.83 & 9.9e-3 & 18.19 & 7.8e-4 & 23.81 & 2.2e-4 & \textbf{28.86}& \textbf{6.9e-5} \\
            & source2 & 0.36 & 9.5e-4 & 0.76 & 8.7e-4 & -2.63 & 1.0e-3 & -0.11 & 9.3e-4 & -4.29 & 6.0e-4 & 4.03 & 3.8e-4 & \textbf{14.25} & \textbf{3.7e-5} \\
            & source3 & -13.79 & 4.0e-4 & -19.95 & 4.0e-4 & -5.59 & 4.6e-4 & -15.76 & 3.9e-4 & -7.26 & 3.2e-4 & 8.9 & 5.3e-5 & \textbf{14.7} & \textbf{3.3e-5}\\
    \hline \\[-2ex]
    Syn. MSig5   & source1 & 2.11 & 1.6e-2 & 15.53 & 1.1e-3 & -4.31 & 2.6e-2 & 1.26 & 1.1e-2 & 18.81 & 5.2e-4 & 19.26 & 4.2e-4 & \textbf{23.97} & \textbf{1.4e-4}\\
            & source2 & -5.27 & 7.4e-4 & 1.02 & 7.0e-4 & -5.64 & 7.2e-4 & -0.05 & 7.3e-4 & -4.42 & 4.3e-4 & 1.27 & 5.5e-4 & \textbf{14.48} & \textbf{2.6e-5} \\
            & source3 & -18.59 & 1.2e-4 & 3.01 & 1.1e-4 & -10.47 & 1.2e-4 & -11.59 & 1.2e-4 & -7.82 & 1.0e-4 & 6.82 & 2.7e-5 & \textbf{15.06} & \textbf{5.1e-6} \\
    \hline \\[-2ex]
           & Average & 0.10 & 9.5e-4 & 8.69 & 5.0e-4 & -4.84 & 1.4e-3 & 1.49 & 6.7e-4 & 11.86 & 3.2e-4 & 18.56 & 2.1e-4 & \textbf{20.88} & \textbf{3.6e-5} \\
    \hline
\end{tabular}}
\end{center}
 \vspace{-5pt}
\end{table*}

\textbf{Metrics:} We present Signal to Distortion Ratio (SDR) and Mean Squared Error (MSE) for each separated source from the five input mixed signals. For averaging MSE values, we employ geometric averaging, whereas for SDR averaging, we use arithmetic averaging in their original linear scale.

\textbf{Experiment Setup:} The spectrograms are generated with window and stride of $60 s$ and $15 s$. Our study shows that an increased time dilation patrameter can improve the performance for extracting sources with longer masked sections. Commonly, the masked sections are longer when targeting sources with higher fundamental frequency than other sources in the mix. This is expected since the stride between harmonics of low-frequency sources are smaller and, therefore, the masked area on the spectrogram is larger. We employed a dilation of either $13$ or $15$, making the decision on a case-by-case basis according to specific masking situation.

Certain conventional signal separation algorithms, such as Independent Component Analysis (ICA) and Principal Component Analysis (PCA), do not accommodate single-detector signal separation and we excluded them from our comparison study. We compare against six previous methods, EMD \cite{EMD}, VMD \cite{VMD}, NMF \cite{NMF}, REPET and REPET-Extended \cite{REPET}, and spectral msking \cite{spectralmasking}. The comparison results is presented in Table \ref{table:finalres}. These results are all calculated on band-pass filtered mixed signals (MSig) between $[0Hz,12Hz]$.

\textbf{Discussion:} The results in Table \ref{table:finalres} show that the DHF method achieves an approximate 26\% ($2.3db$) improvement in Signal-to-Distortion Ratio (SDR) and 80\% reduction in Mean Squared Error (MSE) on average, compared to the best previous signal separation algorithms. Particularly for sources with less than $\times0.1$ amplitude of the dominant source (such as mixed signal 3 source 2, mixed signal 4 source 3, and mixed signal 5 source 3), DHF demonstrates remarkable effectiveness. For these three low-power scenarios, our method shows an average SDR improvement of $7.2db$ and a 92\% average MSE improvement over the best prior methods. In Figure \ref{fig:wavesep} part a, we further analyze the SDR improvements achieved with DHF compared to the highest SDRs of previous methods. The findings indicate that DHF fills a critical gap in existing methods, particularly excelling in scenarios where others falter. We employ a heatmap to illustrate the relationship between the Masked Energy Ratio (defined as the percentage of masked target energy to overall masked energy per signal separation round) and DHF's superior performance. Previous methods tend to struggle with low masked energy ratios, attempting to isolate low-energy signals from strong overlapping interference, a challenge where DHF notably outperforms its predecessors. Figure \ref{fig:wavesep} part b shows an example separated waveform after three rounds of DHF applied on a mixed signal with three sources.

\subsection{Signal Separation Results: \emph{In Vivo} SpO2 Estimation}

\begin{figure}[t]
\vspace{-5pt}
 \hfill\begin{minipage}{.5\textwidth}\centering
  \includegraphics[width=.39\paperwidth]{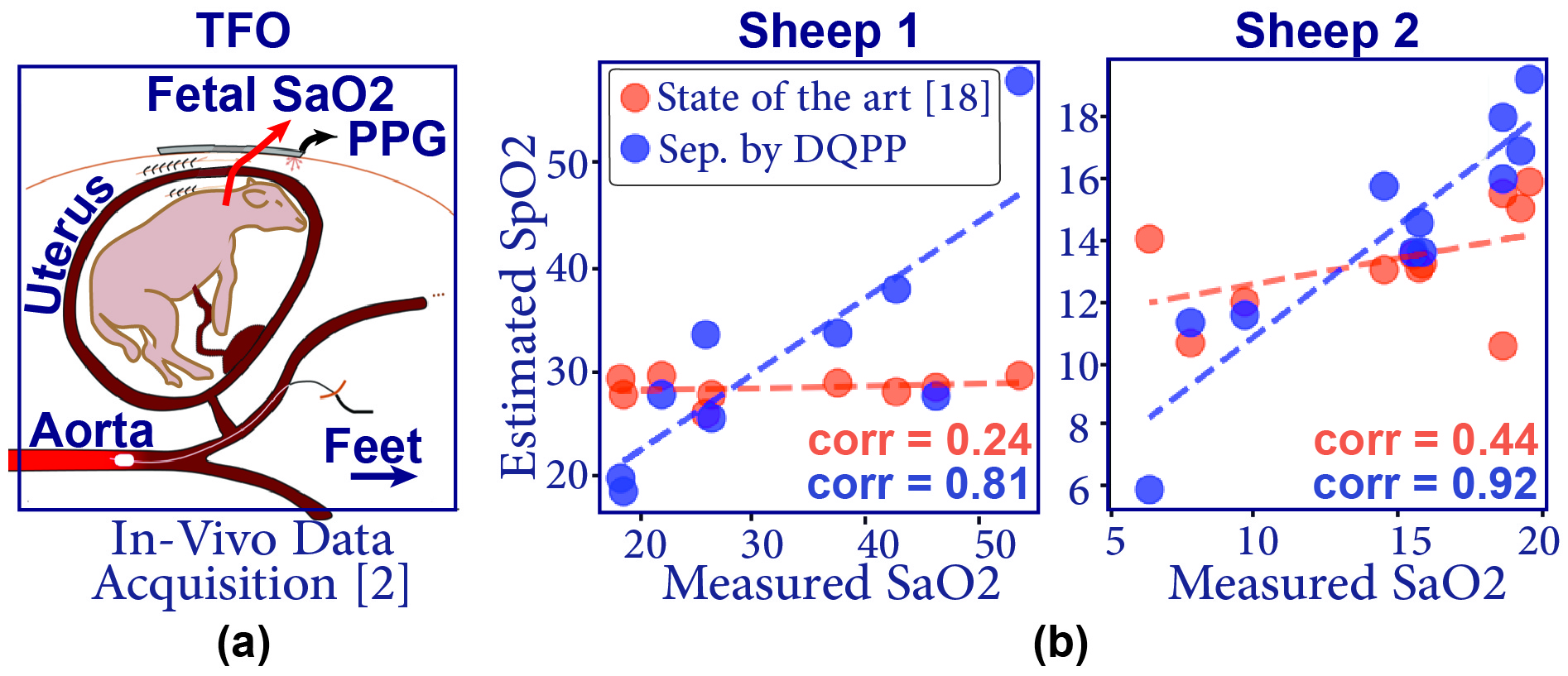}
  \caption{a) TFO data aquisition method \cite{fong2020design}. b) comparison of fetal signal separation and SpO2 estimation using DHF and the state of the art \cite{vali2021estimation}. }
  \label{fig:spo2}
 \end{minipage}
 \vspace{-5pt}
\end{figure}

We study the performance of our approach on fetal SpO2 estimation in an \emph{in vivo} dataset of TFO from two pregnant ewes, sourced from previous studies \cite{vali2021estimation, fong2020design}. This dataset consists of 40 minutes of continuous mixed PPG signals at $740nm$ and $850nm$ wavelengths gatherd from pregnant ewe's abdomen and ground-truth fetal blood oxygen saturation readings, i.e. SaO2, measured from blood-draws with the time distance of 2.5, 5, and 10 minutes. Figure \ref{fig:spo2} part a depicts the TFO data acquisition method used for the adopted dataset. Since the ground-truth fetal PPG signal is not accessible, we report the correlation of SpO2 estimation with measured SaO2 readings when the fetal signal is separated using spectral masking and DHF method. We have compared against spectral masking since it has shown the best performance among previous works.

\begin{figure}[t]
\vspace{-5pt}
 \hfill\begin{minipage}{.5\textwidth}\centering
  \includegraphics[width=.4\paperwidth]{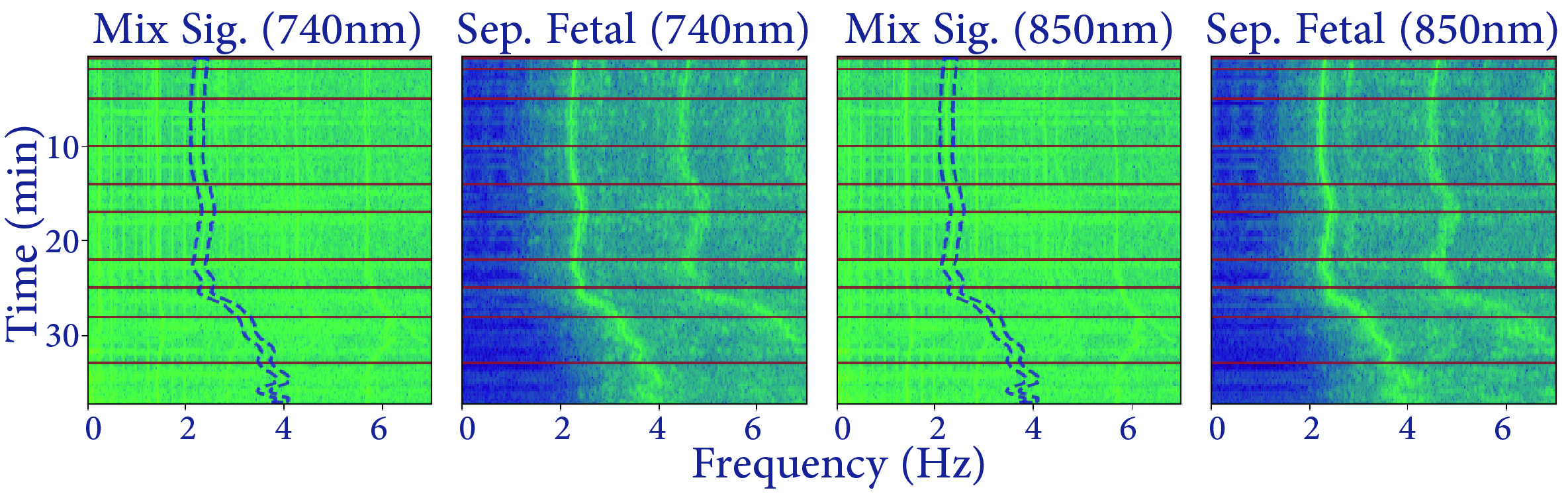}
  \caption{\emph{In vivo} DHF fetal signal separation for sheep 2. Red horizontal lines mark the blood draw timestamps and the dashed-line box highlights the lamb's fetal signal before separation.}
 \label{fig:sheepspect}
 \end{minipage}
 \vspace{-5pt}
\end{figure}

\textbf{SpO2 Estimation:} We use the proposed method in \cite{vali2021estimation} for non-invasive estimation of fetal blood oxygen saturation through pulse oximetery (SpO2). Equation \ref{eq:regression} presents the linear regression method for SpO2 estimation where Y, k, and R are SaO2 measurement vector, a constant regularizing term, and the modulation ratio vector respectively and $w_0$ and $w_1$ are the learned parameters. Equation \ref{eq:Ratio} explains the modulation ratio calculation method which is a classic parameter in pulse oximetry \cite{nitzan2014pulse}. AC and DC represent the dynamic and static portions of the PPG signal at each wavelength. In this work, similar to \cite{vali2021estimation}, R values are averaged over a 2.5 minute window centered around each blood draw time stamp.


\begin{align}
 \vspace{-5pt}
    \label{eq:regression}
    &Y'=\frac{1}{Y+k}=w_0+w_1R, \;\;\; k=1.885 \\
    \label{eq:Ratio}
    &R=\frac{(AC/DC)^{\lambda _1}}{(AC/DC)^{\lambda _2}}
\end{align}

\textbf{Discussion:} Figure \ref{fig:spo2} part b presents the SpO2 estimation performance for both sheeps, comparing the results when the separated fetal signal is obtained through spectral masking (similar to \cite{vali2021estimation}) and DHF. Comparing the correlation between SpO2 estimations and SaO2 readings, our method improves the correlation from 0.24 to 0.81 and from 0.44 to 0.92 in sheep 1 and sheep 2, respectively (80.5\% average error improvement from ideal correlation of 1 between SpO2 estimations and SaO2 readings). Figure \ref{fig:sheepspect} presents the measured mixed signal spectrograms for sheep 2 at 740nm and 850 nm and the separated fetal signal at each wavelength using the DHF method.


\section{Conclusion}

Limitations in dataset capacity and quantity, present in many quasi-periodic signal separation applications in wearable systems, have hindered their success despite their significant potential. Incorporation of application-specific prior knowledge can enhance signal separation performance with limited data. In this work, we use the source frequencies (captured through additional sensors or posterior analytical methods) to, first, mask the undesired frequencies for a specific target source, and second, transform the time-frequency representation to create a stronger alignment between the target source and implicit deep priors built into the structure of our proposed neural network. We showcase our method in the TFO application, using both synthesized and \emph{in vivo} data, where it shows significant improvement compared to the state of the art.

\vspace{-3pt}


\end{document}